\documentstyle[12pt]{article}

\newcommand{\sect}[1]{\setcounter{equation}{0}\section{#1}}

\newfont{\frak}{eufm10 scaled\magstep1}
\newfont{\extra}{msbm10 scaled\magstep1}

\def\be{\begin{equation}}
\def\ee{\end{equation}}
\def\bea{\begin{eqnarray}}
\def\eea{\end{eqnarray}}

\def\om{\omega}
\begin{document}

\begin{center} 
{\LARGE{\bf{Quantum Algebras \\  and \\[0.45cm] Quantum Physics}}}
\end{center}

\bigskip\bigskip

\begin{center} 
E. Celeghini $^1$  and M.A. del Olmo $^2$ 
\end{center}

\begin{center} 
$^1${\sl Departimento di Fisica, Universit\'a  di Firenze and INFN--Sezione di 
Firenze \\ 
I50019 Sesto Fiorentino,  Firenze, Italy}\\ 
\medskip

$^2${\sl Departamento de F\'{\i}sica Te\'orica, Universidad de Valladolid, \\
E-47011, Valladolid, Spain.}\\ 
\medskip

{e-mail:celeghini@fi.infn.it, olmo@fta.uva.es} 
\end{center}

\vskip 1.5cm
\centerline{\today}
\vskip 1.5cm

\bigskip

\begin{abstract} 
In Quantum Mechanics operators must be hermitian and, in a direct product space, symmetric. These
properties are saved by Lie algebra operators  but not by those of quantum algebras. A possible 
correspondence between observables and quantum algebra operators  is suggested by extending the
definition of matrix elements of a physical observable, including the eventual projection on the
appropriate symmetric space.   This allows to build in the Lie space of representations  
one-parameter families of operators belonging to the enveloping Lie algebra that satisfy an
approximate symmetry and have the properties required by physics. 
\end{abstract}
\vskip 1cm

\vskip 0.4cm

\vfill
\eject
\sect{Introduction\label{introduccion}}

Quantum groups arose in the work of the Leningrad school related to the inverse
scattering method \cite{Fad}. Their interest in mathematics is indisputable, and  their
 physical applications cover integrable models, quantum conformal field theories, quantum
field theories, quantum gravity, spin chains, etc. However,  our hopes to rewrite all countless
applications of Lie algebras, with a free parameter inside, has had a limited success up to now.

The reasons  are essentially two. First of all, the essential
 point of applications of  Lie--Hopf algebras to quantum physics is the  one-to-one correspondence
among physical  observables from one side and hermitian operators on a Hilbert space from the 
other. In the case of standard
deformations \cite{Dri,Jim}  such correspondence cannot be extended to quantum
algebras for all values of the deformation parameter $q \in {\cal C}$.
As it is well known, we have to require   $q \in {\cal R}$ or $|q|=1$ to obtain hermitian irreducible
representations. For non-standard quantum algebras \cite{ohn} the situation is worse since raising
and lowering operators have a completely different behaviour for any value of $q$.

The  second reason that stops applications of quantum algebras  to physics is related with  the
concept of composed system: such an object is nothing else that the set of two (or more) sub-systems
that, in some  approximation, can be considered as independent. The fundamental  assumption is
that the Hilbert space is the direct product of  those of the  elementary systems and the
interaction Hamiltonian, when not  disregarded, is such that it does not change this basic structure
 modifying only  the transition matrix elements. 
Physics is indeed described in  direct product spaces  
---at least  we have to pick up the observed system from the rest of the laboratory---
and the scheme must be such that the observables of the composed systems are determined  by the
observables of their  constituents.
In physics all systems have bosonic or fermionic behaviour, i.e., they 
exhibit well defined properties under interchange of their identical constituents. This property is
obviously translated on the direct product space and cannot be modified by the
operators working on it; so, they must be symmetric.  For non-identical constituents this property of
symmetry must be also preserved since physics is independent of the order taken in the direct product
of their wavefunctions.

When  we have a
Lie symmetry, observables are additive (as for the angular  momentum  described in $su(2)$, that for
two particles is simply the sum of the two angular  momenta) and, thus,  symmetric.  For quantum
algebra operators  this property of symmetry is  not verified since  global observables are 
determined by the coalgebra, which is never symmetric (and seldom hermitian) whatever kind of
deformation (standard or non-standard) and value of the deformation parameter are considered.

The point is that both, physics and mathematics, give strong prescriptions on operators structure and
these prescriptions seem to be in contradiction. It looks that nobody can hope to find a solution of
this consistency problem  inside the well established rules of quantum mechanics or quantum algebras
where not contrastable results forbid any possibility.
The only way is to  work on the correlation between quantum observables and quantum algebra
operators.

In other words, the application of Lie algebras to physics is based on two invariances: the first one
is the invariance of the roots of a Lie algebra under the Weyl group which originates
the symmetry between raising and lowering operators. And the second one is the invariance of the
operators acting on ${\cal H}^{\otimes n}$ under the permutation group $S_n$, i.e., they carry the
trivial representation of  $S_n$. 
Quantum algebra operators have not these symmetries and, for this reason they cannot,
in a direct way,  describe physical observables. To avoid these difficulties, we shall consider the
projection  of  the quantum algebra operators on  an appropriate  space in order to define
suitable physical matrix elements.

To illustrate our approach we develop in a detailed way  the simple case of the standard and
non-standard deformations of $su(2)$ because of the physical relevance of their applications and their
computational simplicity. The generalization to higher dimension algebras  is only a technical matter.

\sect{Standard $su_q(2)$}\label{suq2standard} 
A quantum algebra, like $su_q(2)$, is characterized by two algebraic structures: one at the
level of the Lie algebra but with `deformed'  commutators and a second one at the level of the
coalgebra \cite{CP}. 

Let $H,\ X_\pm (=X_1\pm i X_2)$ be the generators of  $su_q(2)$. The deformed 
commutators are
\be\label{conmutadorsu2q}
[H,X_\pm]=\pm 2 X_\pm,\qquad [X_+,X_-]= \frac {\sinh z H}{\sinh z} ,
\ee
with $z=\log q$. Note that when $z\to 0$ we recover $su(2)$.

The $(2j+1)$--dimensional irreducible representations of $su_q(2)$, $D^q_j$, are given by   
\begin{eqnarray}\label{one}
H\;|z,j,m\rangle &=& m \;|z,j,m\rangle, \nonumber \\[0.2cm]
{\cal C}_q \;|z,j,m\rangle &=& [j]_q \;[j+1]_q \; |z,j,m\rangle ,\\[0.2cm]
X_\pm \; |z,j,m\rangle &=& \sqrt{[j\mp m]_q  \; [j\pm m+1]_q} \; |z,j,m\pm 1\rangle , \nonumber
\end{eqnarray}
where $[n]_q={\sinh (z n)}/{\sinh z }$, $2j \in Z^{\geq 0},\ m=-j,-j+1,\dots, j$, and ${\cal
C}_q$ is the deformed Casimir operator  
\be\label{casimir}
{\cal C}_q \equiv X_- X_+ + [H]_q  \; [H+1]_q.
\ee 

At the level of the irreducible representations of $su_q(2)$ both for $z$ real
or imaginary (with $|z|/\pi$ rational and irrational, i.e. with $q$ root of unity or not) the matrix
representations of the generators
$H,\ \ X_1$ and $X_{2}$ as well as the Casimir (\ref{casimir})
 are hermitian. Indeed, in both  cases the scheme is
similar to the nondeformed case, and 
a complete set of commuting observables is composed
by $H$ and ${\cal C}_q$.
Since matrix elements depends on $z^2$, which is real in both cases, one has
\begin{eqnarray}\label{valoresperadoX}
\overline{\langle z,j,n| X_\pm |z,j,m\rangle}= 
\langle z,j,m| X_\mp |z,j,n\rangle ,
\end{eqnarray}
and the usual scalar product $\langle z,j',n |z,j,m\rangle =  \delta_{j',j}\; \delta_{n,m}$ 
is sufficient to define a $*$--representation \cite{CP} with $H,\
X_1$ and $X_2$  hermitian operators and $X_\pm^\dagger =X_\mp$.

We mentioned in the introduction  that most of the difficulties appear when 
composed systems are considered. The structure of Hopf algebra, characteristic of a quantum algebra,
includes in a natural way the composed systems in the  coalgebra.  As it is well known,
the coalgebra is basically determined by the coproduct $\Delta$. For $su_q(2)$ we have 
\begin{eqnarray} \label{coproducto}
\Delta H= H \otimes 1 + 1 \otimes H\;, \qquad
\Delta X_\pm = X_\pm \otimes e^{\frac{z}{2}  H} + e^{-\frac{z}{2} H} \otimes X_\pm  . 
\end{eqnarray}
Note that  $\Delta H$ remains additive and symmetric like in the nondeformed case, while the
expressions of 
$\Delta X_\pm$, imposed by  the commutation relations (\ref{conmutadorsu2q}) on the composed
systems (i.e. $\Delta [\cdot , \cdot]=[\Delta\cdot , \Delta\cdot ]$) become deformed.  As we stressed
before,  physical requirements imply that the operators of a composed system  are  hermitian and
symmetric.  In the present case for
$z$ real  they are hermitian but non symmetric as it is obvious by inspection
of expression  (\ref{coproducto}).
For $z$ imaginary the `naive adjoint'  could look physically acceptable, as the 
change of two component spaces correspond to turn clockwise or anticlockwise in the complex plane. 
However, this is not more true for systems with more that two components.

Thus, also our pragmatic approach cannot escape to  the result
established in general
\cite{CP}: there is an involution for  quantum algebras such that $\Delta  X_1$ and $\Delta  X_2$ 
are hermitian for  $q$ real  only.
However,  this involution is unsatisfactory for a physicist since does not preserve the symmetry
between the factor spaces.

In order to restore this symmetry, let us observe that  the Lie algebra
elements $\Delta(X)= X \otimes 1 + 1 \otimes X$ (X is called primitive element)
carry the trivial representation of the symmetry group $S_2$, i.e. 
\be\label{trivialrepresentation}
\sigma \Delta(X)\sigma^{-1}=\Delta(X),
\ee
where $\sigma$ is the permutation operator ($\sigma (a\otimes b) =b\otimes a$). 
This property (\ref{trivialrepresentation}) is essential for  defining a one-to-one correspondence
between  operators and physical observables. 
The crucial point is to improve  (\ref{trivialrepresentation}) in the deformed
case. 

We propose, thus,  a new definition  of the  matrix elements of an operator $O$ for two
 particles
\be\label{valoresperado2}
\langle \phi| O|\psi\rangle_{\rm phys} :=
\langle \phi|\frac 12 (O + \sigma O \sigma^{-1})|\psi\rangle.
\ee
Notice that if the operator is symmetric this expression is equivalent to the usual one.  We
will denote in general 
\be
\langle \phi| {\widetilde O}|\psi\rangle \equiv\langle \phi| O|\psi\rangle_{\rm phys} ,
\ee
where
\be{\widetilde O} :=\frac 1{2!}\sum_{\sigma \in S_2} \sigma O \sigma^{-1}.
\ee
It is worthy to note that ${\widetilde O}$  carries
the trivial representation of $S_2$ as required and that, because we are projecting on the symmetric
subspace, the physical matrix element of the product $O_1O_2$ is related to $\widetilde{O_1O_2}$ and
not to $\tilde{O}_1\tilde{O}_2$. 

For systems with more that two `particles' the coproduct of high order is build by iteration 
of the coproduct. So
\be
\Delta^{(3)} : A \to A\otimes A\otimes A
\ee
is defined by 
\be
\Delta^{(3)} := ({\rm id} \otimes \Delta^{(2)} ) \circ \Delta^{(2)} =(\Delta^{(2)} 
\otimes  {\rm id} ) \circ \Delta^{(2)},
\ee
where $\Delta^{(2)}=\Delta$. Following this iteration procedure we can obtain $\Delta^{(n)}$
\cite{angel} 
\be
\Delta^{(n)} := ({\rm id} \otimes \Delta^{(n-1)} ) \circ \Delta^{(2)} =(\Delta^{(n-1)} 
\otimes  {\rm id} ) \circ \Delta^{(2)}.
\ee

In general, for a system of $n$   particles  we 
consider the operator 
\be\label{valoresperadon}
{\tilde  O^{(n)}} :=\frac 1{n!}\sum_{\sigma \in S_n} \sigma  O^{(n)}\sigma^{-1},
\ee
which commutes with any permutation of $S_n$ as it is easy to see using the rearrangement lemma.
Remark that this definition  is consistent  with the
usual one for symmetric operators.

Returning to the case of $n=2$ we obtain from (\ref{valoresperado2}) that
\be\label{deltados}
\widetilde{\Delta H} = H \otimes 1 + 1 \otimes  H\qquad
\widetilde{\Delta X_\pm} = X_\pm \otimes \cosh(\frac{z}{2}  H) + 
\cosh(\frac{z}{2}  H) \otimes X_\pm  .  
\ee

For $n=3$ we get from (\ref{valoresperadon}) that
\be\begin{array}{lll}\label{delaathree}
\widetilde{\Delta^{(3)} H} &=& H \otimes 1 \otimes 1 + 1 \otimes  H\otimes 1 +1 \otimes 1 \otimes
H\\[0.2cm]
\widetilde{\Delta^{(3)} X_\pm} &=&\frac 13\{ X_\pm \otimes [ 2  \cosh(\frac{z}{2}  H)\otimes
\cosh(\frac{z}{2}  H)  + \cosh(1\otimes \frac{z}{2}  H +\frac{z}{2}  H \otimes 1)] \\[0.1cm]    
& &\ \ \  + [1 \otimes  X_\pm \otimes 1] [ 2  \cosh(\frac{z}{2}  H)\otimes 1\otimes 
\cosh(\frac{z}{2}  H)\\[0.1cm] 
 & & \qquad\qquad\qquad\qquad \ + \cosh(1\otimes 1 \otimes  \frac{z}{2}  H +\frac{z}{2}  H
\otimes 1 \otimes 1)]\\[0.1cm]  
& &\ \ \ + [ 2  \cosh(\frac{z}{2}  H)\otimes \cosh(\frac{z}{2}  H) 
+ \cosh(1\otimes \frac{z}{2}  H +\frac{z}{2}  H \otimes 1)]  \otimes X_\pm  \}.
\end{array}\ee
Note that expressions (\ref{deltados}) and (\ref{delaathree}) satisfies all the physical requirements
for $z$ real as well as imaginary.

\sect{Non-standard $su_\om (2)$}\label{suq2nonstandard}
The non-standard quantum algebra $su_\om(2)$ \cite{ohn} has the following Hopf algebra structure:
deformed commutators
\be\label{nonstandarcomm}
\begin{array}{lll}
[H,X_+]&=& \frac 2{\om}\ {\sinh{\om X_+}},\\[0.2cm]
 [H,X_-]&=& - X_- (\cosh{\om X_+}) -  (\cosh{\om X_+}) X_-,\\[0.2cm] 
 [X_+,X_-]&=&H ;
\end{array}\ee 
and coalgebra
\be\begin{array}{lll}
\Delta H &=& H\otimes e^{\om X_+}+ e^{-\om X_+}\otimes H, \\[0.2cm]
\Delta X_+ &=& X_+\otimes 1+ 1\otimes X_+, \\[0.2cm]
\Delta X_- &=& X_-\otimes e^{\om X_+}+ e^{-\om X_+}\otimes X_- . 
\end{array}\ee
Notice the different roles played by $X_+$ and $X_-$ in comparison with the standard
deformation case. That requires to `symmetrize' also the irreducible representations.

The fundamental representation ($j=1/2$) is always for quantum algebras  like in the
non-deformed case.  In Ref. \cite{dobrev}  a  3--dimensional ($j=1$)
irreducible representation of $su_\om(2)$  is presented,  and in Ref. \cite{ACC}  
representations for $j=1$ and
$j=3/2$  are displayed with the matrix associated to $H$  diagonal.  Here we consider 
equivalent representations for $j=1$ and $j=3/2$ such that, in the limit of $\om$ going to zero,  
the usual representations of $su(2)$ are recovered. 

For $j=1$ the matrix representation for the generators is
\be\label{3representacion}
H=\left(\begin{array}{cccc}
2&0&0\\ 0&0&0\\ 0&0&-2
\end{array}\right), \quad
X_+=\left(\begin{array}{cccc}
0&\sqrt{2}&0\\ 0&0&\sqrt{2}\\ 0&0&0
\end{array}\right), \quad
X_-=\left(\begin{array}{cccc}
0& -\frac{\om ^2}{2\sqrt{2}} &0\\ \sqrt{2}&0&-\frac{\om ^2}{2\sqrt{2}}\\ 0&\sqrt{2}&0
\end{array}\right);
\ee
and in  the case of $j=3/2$
\be\begin{array}{c}\label{4representacion}
 H=\left(\begin{array}{ccccc}
3&0&0&0\\ 0&1&0&0\\ 0&0&-1&0\\0&0&0&-3
\end{array}\right), \qquad 
X_+=\left(\begin{array}{ccccc}
0&\sqrt{3}&0&\frac{\om ^2}{2}\\ 0& 0&2&0\\  0& 0&0&\sqrt{3}\\  0& 0&0&0\\ 
\end{array}\right), \\ \\
 X_-=\left(\begin{array}{ccccc} 
0& -\frac{\sqrt{3} \om ^2}{2} &0&\frac{3 \om ^4}{8}\\ \sqrt{3}&0&-\frac{3 \om ^2}{2}&0\\
0&2&0&-\frac{\sqrt{3}\om ^2}{2}\\ 0& 0&\sqrt{3}&0
\end{array}\right) .
\end{array}\ee
Representations for higher dimensions are similar.

Physical hermiticity compels us to restore the symmetry between raising and lowering operators.
To achieve it,  let us consider the quantum algebra, ${su}'_{\om}(2)$,  isomorphic to
${su}_{\om}(2)$, obtained by the  transformation
\be
H\to H'=H^\dagger,\qquad X_\pm\to  X'_\pm=X_\mp^\dagger .
\ee
 Let
${\cal H}_j$ and ${\cal H}'_j$ be the carrier spaces of the ($2j+1$)--dimensional irreducible
representations of 
$su_{\om}(2)$ and ${su}'_{\om}(2)$), respectively.  The physical matrix elements of an  operator
$O$ are defined by the mean of the matrix elements in each carrier space, i.e.,
\be
\hat O\equiv (j,n|O|j,m) := \frac 12 (\langle j,n|O|j,m\rangle+\langle j,n | O'|j,m\rangle ).
\ee
 Hermiticity requires that the parameter
 $\om$ has to be  only real or imaginary like in the case of  standard deformations. 

In this way, for the  representations of  $su_{\om}(2)$  we obtain
\be\label{3representacionsmean}
\hat H=\frac 12(H+H^\dagger),\qquad  
\hat X_\pm=\frac 12( X_\pm+ X_\mp^\dagger).
\ee
In the representations (\ref{3representacion}) and (\ref{4representacion}) we have considered
$H=H^\dagger$ and, hence,  $\hat H=H$.

For composed systems the `symmetrization' procedure starts from the symmetrization like the standard
case and finish with the above reported `hermitianization'. In particular, for systems with two
particles the procedure is:

{\bf 1).-} To symmetrize ${su}_{\om}(2)$
\be\label{3nonstandardmean}
\begin{array}{lll}
\widetilde{\Delta H} &=& H\otimes {\cosh (\om X_+)}+\cosh (\om X_+)\otimes H ,\\[0.2cm]
\widetilde{\Delta X_+} &=& X_+\otimes 1+ 1\otimes X_+, \\[0.2cm]
\widetilde{\Delta X_-} &=& X_-\otimes \cosh (\om  X_+)+\cosh (\om  X_+) \otimes X_- ,
\end{array}\ee

{\bf 2).-} To symmetrize ${su}'_{\om}(2)$
\be\label{3nonstandardmeanprima}
\begin{array}{lll}
\widetilde{\Delta H'} &=& H\otimes {\cosh (\om X_+^\dagger)}+\cosh (\om X_+^\dagger)\otimes H ,
\\[0.2cm]
\widetilde{\Delta X_+'} &=& X_-^\dagger\otimes \cosh (\om  X_+^\dagger) 
+\cosh(\om  X_+^\dagger) \otimes X_-^\dagger ,\\[0.2cm]
\widetilde{\Delta X_-'} &=& X_+^\dagger+\otimes 1+ 1\otimes X_+^\dagger.
\end{array}\ee

{\bf 3).-}  To consider the mean of both symmetrization results 
\be\label{3nonstandardmeantotal}
\begin{array}{lll}
\widehat{\widetilde{\Delta H'}} &=& H\otimes \frac12 [\cosh (\om X_+) +
\cosh (\om X_+^\dagger )] +\frac12 [\cosh (\om X_+) +\cosh (\om X_+^\dagger )]\otimes H ,
\\[0.2cm]
\widehat{\widetilde{\Delta X_+'}} &=& \frac12 [X_+\otimes 1 
+ X_-^\dagger\otimes \cosh (\om  X_+^\dagger)+ 1\otimes X_+ 
+ \cosh (\om  X_+^\dagger)\otimes X_-^\dagger],
\\[0.2cm]
\widehat{\widetilde{\Delta X_-'}} &=& \frac12[X_+^\dagger\otimes 1+ X_-\otimes \cosh (\om  X_+)+
1\otimes X_+^\dagger + \cosh (\om  X_+)\otimes X_-].
\end{array}\ee
We do not write the expression for the case of three particles since its computation is
straightforward.

\sect{Conclusions}\label{conclusiones}

 We can sum up the problem as follows: in quantum physics all the observables must have real
eingenvalues and, hence, they must be hermitian. Moreover, they must be symmetric since, as we mention
before, physical results are independent of the order in the direct product spaces. 

On the other hand,  quantum algebra 
generators  are seldom hermitian and never symmetric. Thus, there is not a  physical theory  whose
observables belong to a quantum algebra. We propose to construct physical observables
in terms of quantum algebra operators  enlarging the representation space and later projecting on  an
appropriate  subspace.

We have chosen to present our proposal in an informal way. So, we have simply symmetrized and
hermitianized  quantum algebra operators. A formal description would require the introduction of
a density matrix to be inserted in a reducible representation at the moment of evaluating the matrix
elements. It should be  more clear in this case that the algebraic structure of the set  of quantum
algebra operators has not been modified anyway, and that the correct expression for the product of
operators is
$\widetilde{O_1  O_2}$ because the
density matrix is used only in the last step of the procedure.  

We have considered the standard deformation as well as the non-standard one of
$su(2)$. The procedure is  generalizable to any other Lie algebra with one or other deformation, but
also to quantum algebras with hybrid deformations, i.e,  quantum algebras combining both kinds
of deformations \cite{aneva,olmo}. 

Symmetry in real world is always approximate, but we can find in the literature only one procedure for
describing an approximate symmetry: the spontaneous symmetry breaking that is strange in the sense
that  is an exact symmetry but  manifested  as broken.
We propose here an algebraic approach to broken symmetry that is  between the enormous freedom
we have to build  physical quantities in Lie universal enveloping algebras and the too rigid scheme
to identify them with the generators.
One-parameter families inside  Lie universal enveloping algebras are in this way built that go
with continuity from the exact symmetry ($q=1$)  to a catastrophic breaking ($q\to \infty$ or 
in the neighborhood of a root of  unity).

It is interesting to note that, while an exact symmetry implies that physical quantities are additive,
an approximate symmetry seems to describe objects with correlations between their constituents.

\section*{Acknowledgments}
This work has been partially supported by 
DGES of the  Ministerio de Educaci\'on y Cultura of Spain under 
Projects PB98-0360 and the Junta de Castilla y Le\'on (Spain).  The 
visit of E.C. to Valladolid have been  financed by 
the Universidad de Valladolid.



\begin{thebibliography}{99} 

\bibitem{Fad} L.D. Faddev,  in: Recent Advances in Field Theory and
Statistical Mechanics, eds. J.B. Zuber  and R. Stora R  (North-Holland, Amsterdam, 1984). 

\bibitem{Dri} V.G. Drinfeld, Soviet Math. Dokl. {\bf 32} (1985) 254. 

\bibitem{Jim}  M. Jimbo, Lett. Math. Phys. {\bf 10} (1985) 63.

\bibitem{ohn}  Ch. Ohn, Lett. Math. Phys. {\bf 25} (1992) 85.

\bibitem{CP}  V. Chari and  A. Pressley, A Guide to Quantum Groups 
(Cambridge University Press, Cambridge,  1995).

\bibitem{angel} A. Ballesteros and O. Ragnisco, J. Phys. A {\bf 31} (1998) 3791.

\bibitem{dobrev} V.K. Dobrev, Representations of the Jordanian quantum algebra ${\cal U}_h(su(2))$,
preprint IC/96/14. 

\bibitem{ACC} B. Abdesselam, A. Chakrabarti and R. Chakrabarti,  Mod. Phys. Lett. A, {\bf 11}
 (1996) 2883.

\bibitem{aneva}   B.L. Aneva,  D. Arnaudon,  A. Chakrabarti,  V.K. Dobrev and
S.G. Mihov, J. Math. Phys.  {\bf 42} (2001) 1236.


\bibitem{olmo} V.D. Lyakhovsky, A.M. Mirolubov and M.A. del Olmo, J. Phys. A {\bf 34} (2001)
1467.

\end{thebibliography}
\end{document}